\documentclass[11pt]{article}
\usepackage{graphicx}
\usepackage{amsmath}
\usepackage{makeidx}
\usepackage{indentfirst}

\newcounter{resultnum}[section]

\newcounter{conclusionnum}[section]

\newcounter{conditionnum}[section]

\newcounter{conjecturenum}[section]

\newcounter{examplenum}[section]

\newcounter{exercisenum}[section]

\newcounter{lemmanum}[section]

\newcounter{notationnum}[section]

\newcounter{theoremnum}[section]

\newcounter{definitionnum}[section]

\newcounter{corollarynum}[section]

\newcounter{remarknum}[section]

\newcounter{propositionnum}[section]

\newcounter{acknowledgementnum}[section]

\newcounter{algorithmnum}[section]

\newcounter{axiomnum}[section]

\newcounter{casenum}[section]

\newcounter{claimnum}[section]

\newcounter{summarynum}[section]

\newcounter{problemnum}[section]

\begin{document}

\title{Critical Remarks on Finsler Modifications of Gravity and Cosmology by Zhe Chang and Xin Li}
\date{May 17, 2010}
\author{ Sergiu I. Vacaru\thanks{
sergiu.vacaru@uaic.ro, Sergiu.Vacaru@gmail.com;\newline
http://www.scribd.com/people/view/1455460-sergiu } \\
{\quad} \\
{\small {\textsl{\ Science Department, University "Al. I. Cuza" Ia\c si},} }%
\\
{\small {\textsl{\ 54 Lascar Catargi street, 700107, Ia\c si, Romania}} }}
\maketitle

\begin{abstract}
I do not agree with the authors of papers arXiv: 0806.2184 and  0901.1023v1 (published in Phys. Lett., respectively, B668 (2008) 453 and B676 (2009) 173). They consider that \textit{"In Finsler manifold, there exists a unique linear connection - the Chern connection ... It is torsion freeness and
metric compatibility ... "}. There are well known results (for example, presented in monographs by H. Rund and R. Miron and M. Anastasiei) that in Finsler geometry there exist an infinite number of linear connections defined by the same metric structure and that the Chern  and Berwald connections \textbf{are not metric compatible.} For instance, the Chern's
one (being with zero torsion and "weak" compatibility on the base manifold of tangent bundle) is not generally compatible with the metric structure on total space. This results in a number of additional difficulties and sophistication in definition of Finsler spinors and Dirac operators and in
additional problems with further generalizations for quantum gravity and noncommutative/string/brane/gauge theories. I conclude that standard physics theories can be generalized naturally by gravitational and matter field
equations for the Cartan and/or any other Finsler metric compatible connections. This allows us to construct more realistic models of Finsler spacetimes, anisotropic field interactions and cosmology.

\vskip0.1cm

\textbf{Keywords:}\ Finsler geometry and gravity, modified Friedman
cosmology, modified Newtonian dynamics (MOND).

\vskip3pt

PACS:\ 02.40.-k, 04.25.Nx, 04.50.Kd, 95.35.+d, 95.36.+x, 9.80.Jk
\end{abstract}


There is a recent interest for a new physics beyond the Standard Model
related to Finsler theories of curved spacetime and quantum gravity and
possible applications in modern cosmology, see \cite{pap1,pap2,stavr1,stavr2}
and references therein. Such theories constructed on tangent bundles of
spacetime manifolds are positively with local Lorentz violations which may
be related to new directions in particle physics and dark energy and dark
matter models in cosmology. It is very important that researches in particle
physics and cosmology would became more familiar with some important methods
and perspectives and possible applications of Finsler geometry in standard
and non--standard theories of physics (a survey and reviews of main ideas
and results oriented to applications in high energy physics are presented
respectively in Refs. \cite{vrflg,vrfgnsph,vsgg}).

In this letter I comment on some ambiguities existing in the above mentioned
two papers by Zhe Chang and Xin Li, suggest possible solutions of the
Chern--Finsler ''nonmetricity'' problem and speculate about "well defined"
Finsler gravity theories and cosmological models. The goal is also to
present a brief review of the main concepts and results on Finsler gravity
modifications of the Einstein gravity theory. We emphasize the possibility
to model (pseudo) Finsler configurations as exact solutions in general
relativity and analyze the most important consequences for quantum gravity
and applications in cosmology.

\subsection*{1. Problems with the Chern--Finsler nonmetricity}

Just before formula (5) in \cite{pap1}, the authors wrote: ''In Finsler
manifold, there exist a unique linear connection - the Chern connection
[22]. It is torsion freenes and metric--compatibility, ...''.\footnote{%
we wrote [22] which means the corresponding citation in \cite{pap1}; similar
sentences are written also before formula (2) in \cite{pap2}} \ Perhaps,
such a conclusion was drawn from formulas (5) and (9) in \cite{pap1} stating
that the metric compatibility and zero torsion conditions hold for the Chern
connection on the base manifolds of tangent bundles. In general, Finsler
spaces endowed with Chern and/or Berwald connections, and various their
modifications, are with generic nonmetricity (when the metric and connection
structures are not compatible). The geometry and physical properties of such
Finsler--affine (and generalized Lagrange--affine) spaces and nonholonomic
metric--affine gravity theories were studied in details in Part I of book %
\cite{vsgg}.

We emphasize that a Finsler geometry/gravity model can be defined completely
only on the total space of a tangent bundle $TM$ of a manifold $M$ (see, for
instance, \cite{stavr1,stavr2,vsgg,ma1987,ma} and, chronologically, some
most important monographs on Finsler geometry and applications \cite%
{cartan,rund,asanov,matsumoto,bejancu,bcs}). If a (pseudo) Riemannian
geometry on $M$ is determined only by one fundamental geometric object, the
symmetric metric tensor $\mathbf{g},$ a (pseudo) Finsler geometry has to be
constructed from three fundamental geometric objects on total space $TM.$
\footnote{%
Physicists use "pseudo" for metric structure with local signature $\pm$.
Alternatively, a Finsler geometry/gravity model can be constructed on a
nonholonomic manifold $\mathbf{V}$ endowed with a conventional
"horizontal--vertical", ($h-v$), splitting, $T\mathbf{V}=h\mathbf{V}\oplus v%
\mathbf{V}),$ i.e. with a nonholonomic distribution; we follow notations and
results outlined in \cite{vrflg}, see also details and bibliography on the
geometry and applications of nonholonomic manifolds in \cite{bejf,vsgg}.}
For some canonical Finsler space models, all three fundamental geometric
objects are completely defined by a generating (fundamental) Finsler
function $F(x,y)$ subjected to certain homogeneity and other conditions,
with $x$ denoting the set of local coordinates on $M$ (or, alternatively, on
$h\mathbf{V}$) and $y$ denoting the set of ''fiber'' like coordinates in $TM$
(or, alternatively, on $v\mathbf{V}$). Following well defined conventions
for Cartan/Berwald/Chern--Finsler spaces, one generates by $F$ on $TM$:\ 1)
a (Finsler, for instance, Sasaki type) metric $\ ^{F}\mathbf{g=(}h\ ^{F}%
\mathbf{g,}v\ ^{F}\mathbf{g)},$ 2) a nonlinear connection (N--connection) $%
\mathbf{N}$ (associated to a splitting $TTM=hTM\oplus vTM),$ and 3) a
distinguished connection, d--connection $\ ^{F}\mathbf{D=}\left( h\mathbf{D,}%
v\mathbf{D}\right) $ (which is a linear connection adapted to a
N--connection $h$--$v$ splitting, i.e. preserving under parallelism such a
Whitney sum $\oplus$ stated on corresponding tangent spaces).  So, a Finsler
geometry is completely defined by a corresponding set of data $\left( F:\
^{F}\mathbf{g,N,}\ ^{F}\mathbf{D}\right).$

The authors of \cite{pap1,pap2} cited the monograph \cite{bcs} and used the
Chern d--connection \cite{chern}, $\ ^{F}\mathbf{D}=\ ^{Ch}\mathbf{D},$
derived in a unique form to satisfy the conditions: 1) vanishing torsion, $\
^{Ch}\mathbf{T}=0,$ and ''horizontal'' metric compatibility, $h\mathbf{\ }%
^{Ch}\mathbf{D(}h\ ^{F}\mathbf{\mathbf{g})}=0.$  In brief, the Chern
d--connection $\ ^{Ch}\mathbf{D}$ on $TM$ generalizes the Levi--Civita
connection $\nabla $ on $M$ in such a manner that the torsion is ''pumped''
into a ''vertical'' nonmetricity $v\mathbf{Q}\neq 0,$ when (in general, on $%
TM)$ $\mathbf{Q\equiv \ }^{Ch}\mathbf{D}\ ^{F}\mathbf{g}\neq 0.$ Such a
proof exists in details in \cite{bcs}. So, the authors of \cite{pap1,pap2}
were not right stating that $\mathbf{\ }^{Ch}\mathbf{D}$ is ''metric
compatible''. A ''weak'' compatibility exists only on the h--subspace but
not for a general geometric/physical model on $TM.$ It is also not metric
compatible the Berwald model of Finsler geometry with $\ ^{F}\mathbf{D=}\
^{B}\mathbf{D,}$ (label $B$ is from ''Berwald''), see details in Refs. \cite%
{vsgg,ma1987,ma,bcs}.

Alternatively to $\mathbf{\ }^{Ch}\mathbf{D}$ and/or $\ ^{B}\mathbf{D,}$
there is in Finsler geometry a canonical metric compatible d--connection
(called the Cartan d--connection \cite{cartan}), $^{F}\mathbf{D=}\ ^{c}%
\mathbf{D}$, (historically, it was the first one introduced for a completely
defined model of Finsler space). It satisfies the metricity condition on $TM$%
, $\ ^{c}\mathbf{D}\ ^{F}\mathbf{g}=0,$ but has a nontrivial torsion, $%
\mathbf{\ }^{c}\mathbf{T}\neq 0.$ The interesting thing is that $\mathbf{\ }%
^{c}\mathbf{T}$ is induced by h- and v--components of $\ ^{F}\mathbf{g} $
and the coefficients of the canonical Cartan N--connection, $\mathbf{N=\
^{c}N},$ (with an associated canonical semi-spray configuration \cite%
{vsgg,ma1987,ma,bcs}). All mentioned canonical values are determined by a
fundamental Finsler function $F.$ Such a torsion $\mathbf{\ }^{c}\mathbf{T}$
is very different from those in Einstein--Cartan/ gauge / string gravity
when certain additional field equations are used for determining the torsion
components.

In general, there is an infinite number of metric compatible connections in
Finsler and Lagrange geometries (Lagrange geometry is a "nonhomogeneous"
generalization of Finsler geometry, with a nondegenerate fundamental
Lagrangian); see a theorem by R. Miron in \cite{ma1987,ma}, we discuss the
physical implications in \cite{vrflg}. So, the authors of \cite{pap1,pap2}
erred stating that the Chern connection is the ''unique'' one for Finsler
gravity. Here we also emphasize that the nonmetricity of Finsler gravity
models with Chern/ Berwald d--connection, and with any other nonmetric one,
results in more sophisticate "very nonstandard" physical theories. Together
with an unclear status of nonmetricity fields, the metric incompatibility
make more difficult the definition of spinors and conservation laws in
Finsler gravity (there is a series of our works in this direction \cite%
{vsp1,vsp2,vspstav,vspvicol,vspalg}) and does not allow "simple" (super)
string and noncommutative generalizations like we proposed \cite%
{vstr1,vstr2,vspnc,vncg}.\footnote{%
Briefly, we sketch the problem of definition of Clifford structures and
spinors and of conservation laws in Finsler spaces:\ In general relativity,
we have $\gamma _{\mu }\gamma _{\nu }+\gamma _{\nu }\gamma _{\mu }=2g_{\mu
\nu },$ for gamma matrices $\gamma _{\mu },$ when the Levi--Civita
connection $\nabla$\ satisfies the metricity conditions $\nabla _{\alpha
}g_{\mu \nu }=0.$ This allows us to introduce the Dirac operator on
(psedudo) Riemannian spaces, induced by $\nabla _{\alpha }$ and a fixed
tetradic basis and to define the Dirac equations for the Einstein gravity
theory when $\nabla _{\alpha }T^{\mu \nu }=0,$ for any matter fields with
energy--momentum tensor $T^{\mu \nu }.$%
\par
For a Finsler gravity model with nonmetricity, when $^{F}\mathbf{D}\ ^{F}%
\mathbf{\mathbf{g\neq }}0,$ it is a problem to define self--consistent
analogs of gamma matrices and (Finsler-) Dirac operators. Working with
metric compatible d--connections with nonholonomically induced torsions, the
constructions become similar to those in general relativity. Here we also
note that for nonholonomic configurations the Ricci tensor, in general, is
not symmetric and $^{F}\mathbf{D}_{\alpha }\ T^{\mu \nu }\mathbf{\mathbf{%
\neq }}0.$ Both the Finsler--Dirac operators and generalized conservation
laws on Finsler space with metric compatible d--connections, and
corresponding nonintegrable constraints, were constructed/defined in the
works to which this footnote refers.} We studied in details the so--called
Finsler/Lagrange -- affine spaces and gravity models, in general, with
nontrivial torsion and nonmetricity, in Part I of monograph \cite{vsgg} and
discussed possible connections of Finsler geometry and methods to standard
theories of physics in \cite{vrflg,ijgmmp}.

\subsection*{2. Ambiguities with Finsler generalizations of the Einstein
equations}

Before formula (9) in \cite{pap2}, the authors wrote: ''The Ricci tensor on
Finsler manifold was first introduced by Akbar--Zadeh [21]'' (we put in this
letter the reference \cite{akbar} which in that paper is [21]). Perhaps that
was the first attempt to define a Ricci tensor for Finsler spaces with
constant sectional curvature for the Chern d--connection $\mathbf{\ }^{Ch}%
\mathbf{D,}$ but this is not correct for all Finsler geometry/gravity
models. In general, there were considered various types of Ricci type
tensors in Finsler geometries (see, for instance, \cite%
{rund,asanov,ma1987,ma,bejancu,vsgg,vrflg}). There is, for instance, a
discussion on Finsler and Lagrange gravity by S. Ikeda in Appendix to \cite%
{ma1987}. We also analyzed in details such generalized Finsler gravity
theories, with zero and non--zero nonmetricity, for ''standard'' and
''nonstandard'' theories of particle physics and gravity, providing a number
\ of exact solutions and applications (together with generalizations to
non--commutative/string/grave gravity), see reviews \cite%
{vrflg,vrfgnsph,ijgmmp,vsgg}.

We note that in monographs \cite{ma1987,ma} (see also references therein),
the Einstein equations were first time written in a self--consistent form on
vector/tangent bundles for the so--called canonical d--connection/
h--v--connecti\-on (in our works, for instance, \cite{vrflg,ijgmmp} denoted $%
\ \widehat{\mathbf{D}};$ see there further details and local formulas). For
canonical $\mathbf{\ ^{c}N}$ and $\ ^{c}\mathbf{D},$ they contain the
Einstein equations for the Cartan--Finsler gravity model on $TM.$ In brief,
we can say that such a Finsler gravity theory is similar to the Einstein
gravity but the field equations are for $\ ^{c}\mathbf{D}.$ The
corresponding Ricci, $Ric(\widehat{\mathbf{D}}),$ and Einstein, $\mathbf{E}(%
\widehat{\mathbf{D}}),$ tensors are determined not by the Levi--Civita
connection $\nabla $ (which is not a d--connection because it is not adapted
to the N--connection structure) but constructed for $\widehat{\mathbf{D}}.$
In a more particular case, for Finsler and Lagrange spaces, we can consider $%
\widehat{\mathbf{D}}=\ ^{c}\mathbf{D}.$

In Refs. \cite{ma1987,ma}, for gravity theories on (generalized)
Lagrange--Finsler spaces, on $TM,$ and Finsler like configurations modelled
on nonholonomic manifolds (for instance, defined as exact solutions  in
Einstein gravity)\footnote{%
for our purposes, a nonholonomic manifold is a usual (pseudo) Riemannian
one with a prescribed nonholonomic distribution; a Finsler space can be
considered as an example of nonholonomic tangent bundle with a corresponding
nonintegrable (equivalently, nonholonomic/anholonomic) distribution
determined by a N--connection structure, see details in \cite%
{vrflg,vrfgnsph,ijgmmp,vsgg,ma,ma1987}}, there were considered gravitational
\ field equations of type
\begin{equation}
\mathbf{E}(\widehat{\mathbf{D}})=\mathbf{\Upsilon },  \label{deinst}
\end{equation}%
where the Einstein tensor $\mathbf{E}(\widehat{\mathbf{D}})$ and source $%
\mathbf{\Upsilon }$ are constructed from $\widehat{\mathbf{D}},\ ^{F}\mathbf{%
g}$ and Lagrangians for matter fields following the same principles as in
general relativity theory but extended on corresponding vector/tangent
bundles, or nonholonomic manifolds, in N--adapted form (following a
corresponding tensor/form/spinor/variational calculus preserving the h--/
v--splitting). We can consider such $\mathbf{\Upsilon }$ when for $\widehat{%
\mathbf{D}}\rightarrow \nabla $ the equations (\ref{deinst}) transform into
the usual Einstein equations (for different purposes, and generality, we
have to work with arbitrary dimensions), see details in \cite%
{vrflg,ijgmmp,vncg}. We emphasize that
\begin{equation}
\widehat{\mathbf{D}}=\nabla +\widehat{\mathbf{Z}},  \label{dist}
\end{equation}%
when both the linear connections $\widehat{\mathbf{D}}$ and $\nabla $ and
the distortion tensor $\widehat{\mathbf{Z}}$ are defined by the same metric
structure $\mathbf{g}=\ ^{F}\mathbf{g}$ (we can introduce Finsler variables
by certain convenient frame/coordinate transforms).

The most important property of Finsler like theories of gravity is that the
"locally anisotropic" gravitational field equations are formulated not for
the Levi--Civita connection $\nabla $ but for a d--connection $\mathbf{D,}$
or $^{F}\mathbf{D}.$ In explicit form, a "physical" d--connection  has to be
chosen following certain theoretical arguments and compatibility with
experimental data. If $\mathbf{D=}$ $\widehat{\mathbf{D}},$ or $\mathbf{D=}\
^{c}\mathbf{D,}$ we work with metric compatible geometries and more
realistic physical models (admitting fermionic and gauge fields which can be
introduced similarly as in standard particle physics). Even, in general,
\begin{equation}
\widehat{\mathbf{D}}\left( \mathbf{E}(\widehat{\mathbf{D}})\right) =\widehat{%
\mathbf{D}}\mathbf{\Upsilon \neq 0,}  \label{conslaw}
\end{equation}%
contrary to the general relativity theory when
\begin{equation}
\nabla E=\nabla T(matter)=0,  \label{einstcl}
\end{equation}%
we can consider that \ relations of type (\ref{conslaw}) follow from (\ref%
{einstcl}) and usual Bianchi identities for (pseudo) Riemannian spaces; this
is uniquely defined by nonholonomic distortions of type (\ref{dist}). For
Cartan--Finsler spaces, such generalized conservation laws (\ref{conslaw})
are uniquely determined by a fundamental Finsler function $F(x,y)$ via
corresponding fundamental data $\left(\ ^{F}\mathbf{g,N,}\ ^{F}\mathbf{D}%
\right) .$

The solutions for Randers--Finsler gravity and cosmology (with the Chern
d--connection of approximate Berwald type etc) provided in Refs. \cite%
{pap1,pap2} are supposed to solve certain important issues in modified
Einstein/Newton gravity and dark energy and dark matter new physics. The
gravitational field equations considered in those works are for
d--connections which are metric noncompatible and resulting in a number of
conceptual and theoretical problems in defining (for instance) conservation
laws, spinors and Dirac equations, and a less clear status for nonmetricity
fields. Introducing various types of nonmetricity fields we can "suit a lot
of experimental data". Nevertheless, the main question is this: Why in
Finsler gravity theories we should use metric noncompatible connections if
there are various types of metric compatible ones with less problems for
"standard" physics and without strong experimental constraints analyzed in %
\cite{will}? Here we also emphasize that for the Finsler like configurations
derived as exact solutions of Einstein equations in general relativity, i.e.
on nonholonomic pseudo--Riemannian manifolds, the local Lorentz symmetry can
be preserved, see \cite{vrflg,ijgmmp}. Such constructions can be naturally
extended on (co) tangent bundle theories even in such cases we can not avoid
models with broken local symmetries. Our conclusion is that we may construct
more "standard" physical Finsler classical/quantum gravity theories for
metric compatible connections like the Cartan d--connection.

\subsection*{3. On ''well defined'' Finsler gravity theories and
cosmological models}

It seems that Finsler like gravity theories on (co) tangent bundles (with
metrics and connections depending on velocity/momentum type variables) are
natural consequences of all models of quantum gravity, see physical
arguments and a review of recent results provided in \cite{lammer}. A
principle of general covariance coming from the classical version of the
Einstein gravity theory results in very general quantum nonlinear
dispersions of Finsler and non--Finsler type. Perhaps, it is not the case to
postulate from the very beginning that such a generalized Finsler spacetime
is of any special Randers/Berwald/Chern/...-- Finsler type with fixed (in
general, non--quadratic) line elements like that, for instance, taken for
the Very Special Relativity etc.

We argue that the quantum gravity theory is "almost sure" of generalized
Finsler type on a correspondingly quantized (co) tangent bundle which in
certain classical limits is described by nonholonomic gravity configurations
on (pseudo) Riemannian/--Finsler spacetimes and possible observable effects
in modern cosmology and quantum physics. It can be approached following well
defined geometric and physical principles when the concepts of metric,
connection and frame of reference are postulated to be the fundamental ones
(even, in general, as certain quantized fields and/or possibly redefined as
some almost K\"{a}hler/ generalized Poisson structures etc) in any spacetime
geometry and gravity theories:

\begin{enumerate}
\item For general nonlinear non--quadratic line elements, we can consider
generating fundamental Finsler, or Lagrange (on cotangent bundles,
respectively, Cartan, or Hamilton; in general, of higher order, see Refs. %
\cite{vstr2,vsp2,vspstav,vspvicol,m1,m2,vd1,vd2}) functions. Lifts of Sasaki
type, or another ones, allow us to define canonical (Finsler type and
generalizations) metric, $\ ^{F}\mathbf{g,}$ and N--connection, $\ ^{c}%
\mathbf{N,}$ structures.

\item From the class of infinite number of metric compatible and
noncompatible linear (Finsler type, or generalized) connections, we can
always choose/construct, following the so--called Kawaguchi and Miron
processes \cite{ma1987,ma,vsgg,vrflg}, a canonical d--connection $\widehat{%
\mathbf{D}}. $ In particular, we can introduce, for any (pseudo) Finsler
geometry, the Cartan d--connection, $\ ^{c}\mathbf{D,}$ which is metric
compatible and completely defined by $\ ^{F}\mathbf{g}$ and $\ ^{c}\mathbf{N.%
}$ This way we eliminate possible difficulties/sophistication related to the
nonmetricity geometry and fields and may consider, or derive in certain
limits, various types of Finsler--Lagrange (super) string, gauge,
nonholonomic Clifford/spinor, Finsler--affine and/or noncommutative gravity
theories \cite{vsgg,vsp1,vsp2,vspalg,vspnc,vncg,vstr1,vstr2}.

\item The Einstein equations for $\widehat{\mathbf{D}},$ or $\ ^{c}\mathbf{D}
$, (using nonholonomic constraints, we can include  theories with $\nabla
)$, can be solved in very general forms for different models of Einstein and
Finsler gravity, and various noncommutative/ supersymmet\-ric etc
generalizations, following the anholonomic frame deformation method, see
reviews of results in applications in \cite{ijgmmp,vncg,vsgg,vrflg,vexsol}.
From various classes of very general generic off--diagonal solutions (with
metrics which can not be diagonalized by coordinate transforms), we can
chose well defined subclasses having certain physical importance (describing
locally anisotropic black hole/ellipsoid/torus configurations, cosmological
inhomogeneous and locally anisotropic solutions, solitons etc, see various
examples and reviews in \cite{vsingl,vgon,vcosm1,vrflg,vsgg,ijgmmp}). \

\item For applications in modern cosmology, for instance, with the aim to
elaborate realistic Finsler like inflation, dark energy and dark matter
scenarios, it is important to elaborate Finsler generalizations (they are,
in general, with inhomogeneous and anisotropic metrics) of the Friedman and
Robertson--Walker (FRW) models. Such cosmological models should be grounded
on solutions of the (Finsler) Einstein equations (\ref{deinst}) for certain
types of d--connections (with the above mentioned priorities for the metric
compatible ones). If such models may propose certain important ideas and
solutions in modern cosmology, they would serve as explicit criteria for
choosing as the fundamental ones certain examples of Finsler like linear and
nonlinear connections, and relevant anisotropic metric configurations. It is
not the case to postulate from the very beginning any particular cases of
metric ansatz and/or d--connections for Randers/Berwald/Chern/...-- Finsler
spacetime even they may be preferred as some more "simple" generalizations
of the (pseudo) Riemannian spacetimes. Viable gravity theories are with
nonlinear field equations and the exact solutions for cosmology and
astrophysics, in general, are for generic off--diagonal metrics and
generalized connections. A rigorous mathematical approach does not obey
obligatory any original assumptions on parametrization of metrics and
connections and conventional splitting into "holonomic" and "nonholonomic"
variables.

\item There are also two another very important properties of the Cartan
d--connection (which do not exist for the Chern/Berwald and other metric
noncompatible d--connections):

\begin{itemize}
\item $\ ^{c}\mathbf{D}$ and $\ ^{F}\mathbf{g,}$ for a fixed $\ ^{c}\mathbf{%
N,}$ define canonical almost K\"{a}hler models of Finsler--Lagrange,
Hamilton--Cartan, Einstein gravity and various generalizations. Such models
can be quantized applying a nonholonomically generalized Fedosov method \cite%
{vfclfg,vegpla,vanast}, following the A--brane formalism \cite{vbrane}, and
developing a two--connection perturbative approach to the Einstein and gauge
gravity theories \cite{vgauge1,vgauge2}. Such quantum Einstein and/or
Finsler--Lagrange gravity theories can be elaborated to have in certain
quasi--classical limits different terms with locally violated Lorentz
invariance, anomalies, formal renormalization properties etc. It is also
possible to construct models limiting locally relativistic and covariant
theories.

\item Finsler--Lagrange evolutions of geometry/gravity theories \cite%
{vrfrf,vromp}, and various generalizations with nonsymmetric metrics \cite%
{vnsrf}, noncommutative corrections \cite{vspnc} etc, appear naturally if
Ricci flows of (pseudo) Riemannian metrics are subjected to nonholonomic
constraints on evolution equations, see also possible applications in modern
gravity, cosmology and astrophysics \cite{vrf1,vvis1}. They are positively
related via certain nonlinear renorm group flows to fundamental problems in
quantum gravity and ''early'' stage of anisotropic quantum universes.
\end{itemize}
\end{enumerate}

Finally we conclude:\ There are two general classes of Finsler type gravity,
and geometric mechanics, theories with applications in modern physics and
cosmology.  The first class of locally anisotropic gravity theories
originates from E. Cartan works on Finsler geometry, spinors and bundle
spaces.\footnote{%
he was the first who introduced the N--connections, in component form, put
the basis of Einstein--Cartan theories, elaborated the moving frame method
and the geometry of differential equations in the language of Pfaff forms etc%
} Here we cite the monograph \cite{cartan} and further developments in \cite%
{rund,asanov,matsumoto,ma1987,ma,bejancu,bejf,m1,m2,vstr1,vstr2,vncg,vsgg,vrflg}%
. Even in the just mentioned monographs and review papers a number of
geometric and physical constructions and Finsler geometry methods were
considered for both types of metric compatible or noncompatible
d--connections in Finsler spaces, the most related to ''standard physics''
constructions were elaborated for the Cartan and canonical d--connections
which are metric compatible and follow the geometric and physical principles
1-5 mentioned above. Alternatively (the second class of theories),  there
are Finsler geometry and gravity models grounded on the Berwald and Chern
d--connections, see details in \cite{chern,bcs,pap1,pap2}, and a comparative
review of standard and nonstandard physical theories and applications in
Part I of \cite{vsgg} and in Refs. \cite{vrflg,vrfgnsph}.

The key issues which should be solved both theoretically and experimentally
are those if certain fundamental problems in quantum gravity and/or modern
cosmology can be approached following Finsler theories with metric
compatible, or not compatible, d--connections. The recent interest in new
Finsler gravity physics and cosmology was in the bulk oriented to models
both with local Lorentz violations and nonmetricity, like \cite{pap1,pap2}.
It would be a grave error if non--experts in Finsler geometry but physicists
and mathematicians working in gravity and particle physics and/or cosmology
would consider that the Chern d--connection is a ''unique metric compatible
and the best one'' for Finsler like theories. The reality is that only
following approaches with metric compatible connections, like the Cartan
d--connection, we can elaborate physically viable models which are closely
related to standard physics (as we emphasized in our works, see \cite%
{vfclfg,vegpla,vanast,vbrane,vgauge2,vrfrf,vvis1,vexsol,vsingl,vgon,vspnc,vncg,vstr2}
and reviewed in \cite{vrflg,ijgmmp,vsgg}).

\vskip5pt

\textbf{Acknowledgement: } The author is grateful to R. Miron, M. Anastasiei
and P. Stavrinos for important discussions and kind support.

\end{document}